\documentclass[aps,superscriptaddress,preprintnumbers,amsmath,amssymb]{revtex4}

\usepackage{graphicx} 
\usepackage{dcolumn}  
\usepackage{amsmath} 
\usepackage{epsfig}
\usepackage{epstopdf}

\usepackage{color}

\begin{document}

\preprint{\vbox{ \hbox{   }
  		               \hbox{Belle Preprint 2011-1}
  		               \hbox{KEK Preprint 2010-49}
}}

\title{ \quad\\[0.5cm]   \boldmath Measurements of time-dependent $CP$ asymmetries in $B \to D^{*\mp} 
\pi^{\pm}$ decays using a partial reconstruction technique
}

\affiliation{Budker Institute of Nuclear Physics, Novosibirsk}
\affiliation{Faculty of Mathematics and Physics, Charles University, Prague}
\affiliation{University of Cincinnati, Cincinnati, Ohio 45221}
\affiliation{Justus-Liebig-Universit\"at Gie\ss{}en, Gie\ss{}en}
\affiliation{Hanyang University, Seoul}
\affiliation{University of Hawaii, Honolulu, Hawaii 96822}
\affiliation{High Energy Accelerator Research Organization (KEK), Tsukuba}
\affiliation{Indian Institute of Technology Guwahati, Guwahati}
\affiliation{Institute of High Energy Physics, Chinese Academy of Sciences, Beijing}
\affiliation{Institute of High Energy Physics, Vienna}
\affiliation{Institute of High Energy Physics, Protvino}
\affiliation{Institute for Theoretical and Experimental Physics, Moscow}
\affiliation{J. Stefan Institute, Ljubljana}
\affiliation{Institut f\"ur Experimentelle Kernphysik, Karlsruher Institut f\"ur Technologie, Karlsruhe}
\affiliation{Korea Institute of Science and Technology Information, Daejeon}
\affiliation{Korea University, Seoul}
\affiliation{Kyungpook National University, Taegu}
\affiliation{\'Ecole Polytechnique F\'ed\'erale de Lausanne (EPFL), Lausanne}
\affiliation{Faculty of Mathematics and Physics, University of Ljubljana, Ljubljana}
\affiliation{University of Maribor, Maribor}
\affiliation{Max-Planck-Institut f\"ur Physik, M\"unchen}
\affiliation{University of Melbourne, School of Physics, Victoria 3010}
\affiliation{Nagoya University, Nagoya}
\affiliation{Nara Women's University, Nara}
\affiliation{National Central University, Chung-li}
\affiliation{National United University, Miao Li}
\affiliation{Department of Physics, National Taiwan University, Taipei}
\affiliation{H. Niewodniczanski Institute of Nuclear Physics, Krakow}
\affiliation{Nippon Dental University, Niigata}
\affiliation{Niigata University, Niigata}
\affiliation{University of Nova Gorica, Nova Gorica}
\affiliation{Novosibirsk State University, Novosibirsk}
\affiliation{Osaka City University, Osaka}
\affiliation{Panjab University, Chandigarh}
\affiliation{Saga University, Saga}
\affiliation{University of Science and Technology of China, Hefei}
\affiliation{Seoul National University, Seoul}
\affiliation{Sungkyunkwan University, Suwon}
\affiliation{School of Physics, University of Sydney, NSW 2006}
\affiliation{Tata Institute of Fundamental Research, Mumbai}
\affiliation{Excellence Cluster Universe, Technische Universit\"at M\"unchen, Garching}
\affiliation{Toho University, Funabashi}
\affiliation{Tohoku Gakuin University, Tagajo}
\affiliation{Tohoku University, Sendai}
\affiliation{Tokyo Institute of Technology, Tokyo}
\affiliation{Tokyo Metropolitan University, Tokyo}
\affiliation{Tokyo University of Agriculture and Technology, Tokyo}
\affiliation{CNP, Virginia Polytechnic Institute and State University, Blacksburg, Virginia 24061}
\affiliation{Wayne State University, Detroit, Michigan 48202}
\affiliation{Yonsei University, Seoul}
  \author{S.~Bahinipati}\affiliation{University of Cincinnati, Cincinnati, Ohio 45221} 
   \author{K.~Trabelsi}\affiliation{High Energy Accelerator Research Organization (KEK), Tsukuba} 
  \author{K.~Kinoshita}\affiliation{University of Cincinnati, Cincinnati, Ohio 45221} 
  \author{K.~Arinstein}\affiliation{Budker Institute of Nuclear Physics, Novosibirsk}\affiliation{Novosibirsk State University, Novosibirsk} 
  \author{V.~Aulchenko}\affiliation{Budker Institute of Nuclear Physics, Novosibirsk}\affiliation{Novosibirsk State University, Novosibirsk} 
  \author{T.~Aushev}\affiliation{\'Ecole Polytechnique F\'ed\'erale de Lausanne (EPFL), Lausanne}\affiliation{Institute for Theoretical and Experimental Physics, Moscow} 
 
  \author{A.~M.~Bakich}\affiliation{School of Physics, University of Sydney, NSW 2006} 
  \author{V.~Balagura}\affiliation{Institute for Theoretical and Experimental Physics, Moscow} 
  \author{E.~Barberio}\affiliation{University of Melbourne, School of Physics, Victoria 3010} 
  \author{K.~Belous}\affiliation{Institute of High Energy Physics, Protvino} 
% ...
\collaboration{The Belle Collaboration}

\begin{abstract}
We report results on time-dependent $CP$ asymmetries in $B \to D^{*\mp}\pi^{\pm}$ decays based on  a data sample containing 657 x $10^6$ $B\overline{B}$ pairs collected with the Belle detector at the KEKB asymmetric-energy $e^+ e^-$ collider at the $\Upsilon(4S)$ resonance. We use a partial reconstruction technique, wherein signal $B \to D^{*\mp}\pi^{\pm}$ events are identified using information only from the fast pion from the $B$ decay and the slow pion from the subsequent decay of the $D^{*\mp}$, where the former (latter) corresponds to $D^{*+} (D^{*-} )$ final states. We obtain $CP$ violation parameters
  $S^+  = +0.061 \pm 0.018(\mathrm{stat}) \pm 0.012(\mathrm{syst})$ and
  $S^-  = +0.031 \pm 0.019(\mathrm{stat}) \pm  0.015(\mathrm{syst})$. \\
  
  \end{abstract}
\maketitle

\tighten

{\renewcommand{\thefootnote}{\fnsymbol{footnote}}}
\setcounter{footnote}{0}

\begin{figure}[htb]
\begin{minipage}{4.275cm}
\includegraphics[width=4.25cm,clip]{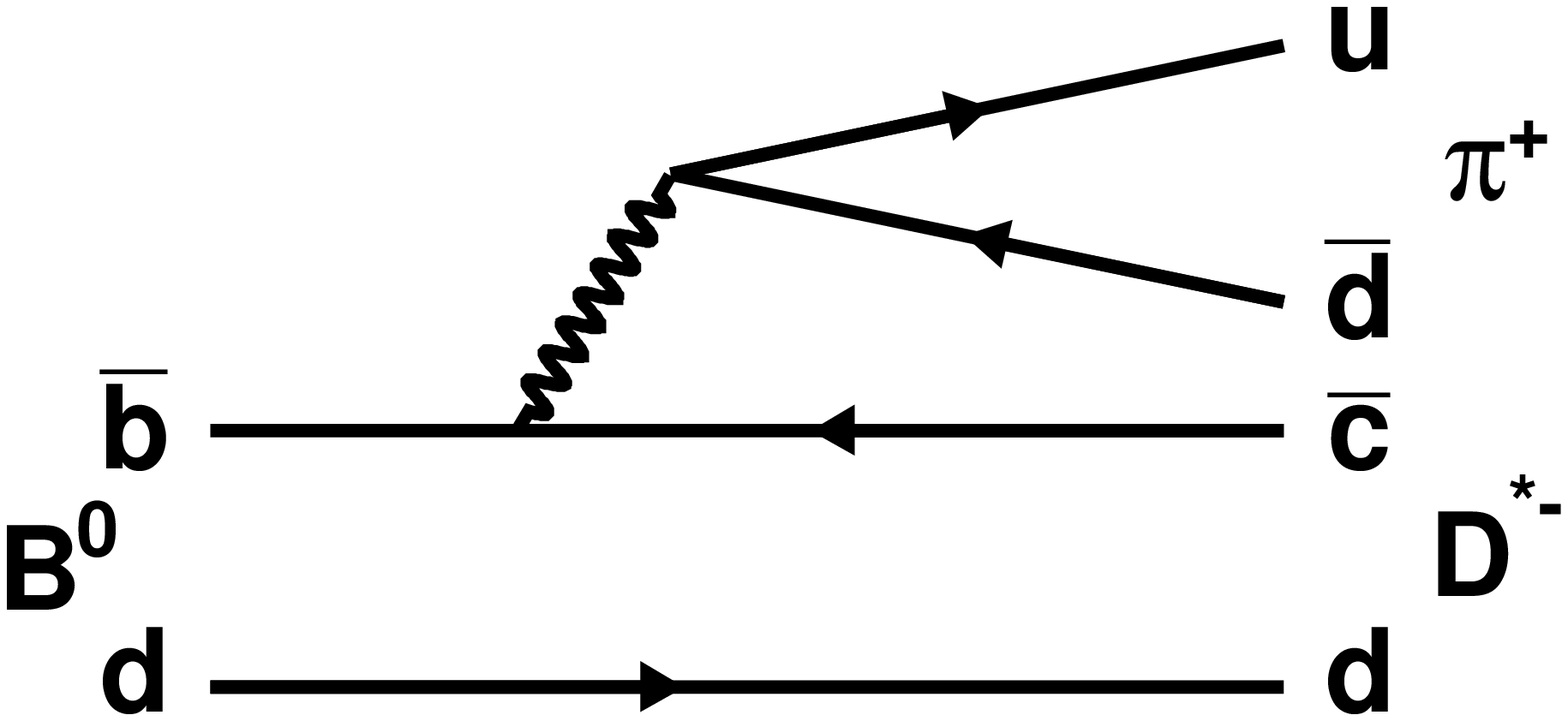}\\
\end{minipage}
\begin{minipage}{4.275cm}
\includegraphics[width=4.25cm,clip]{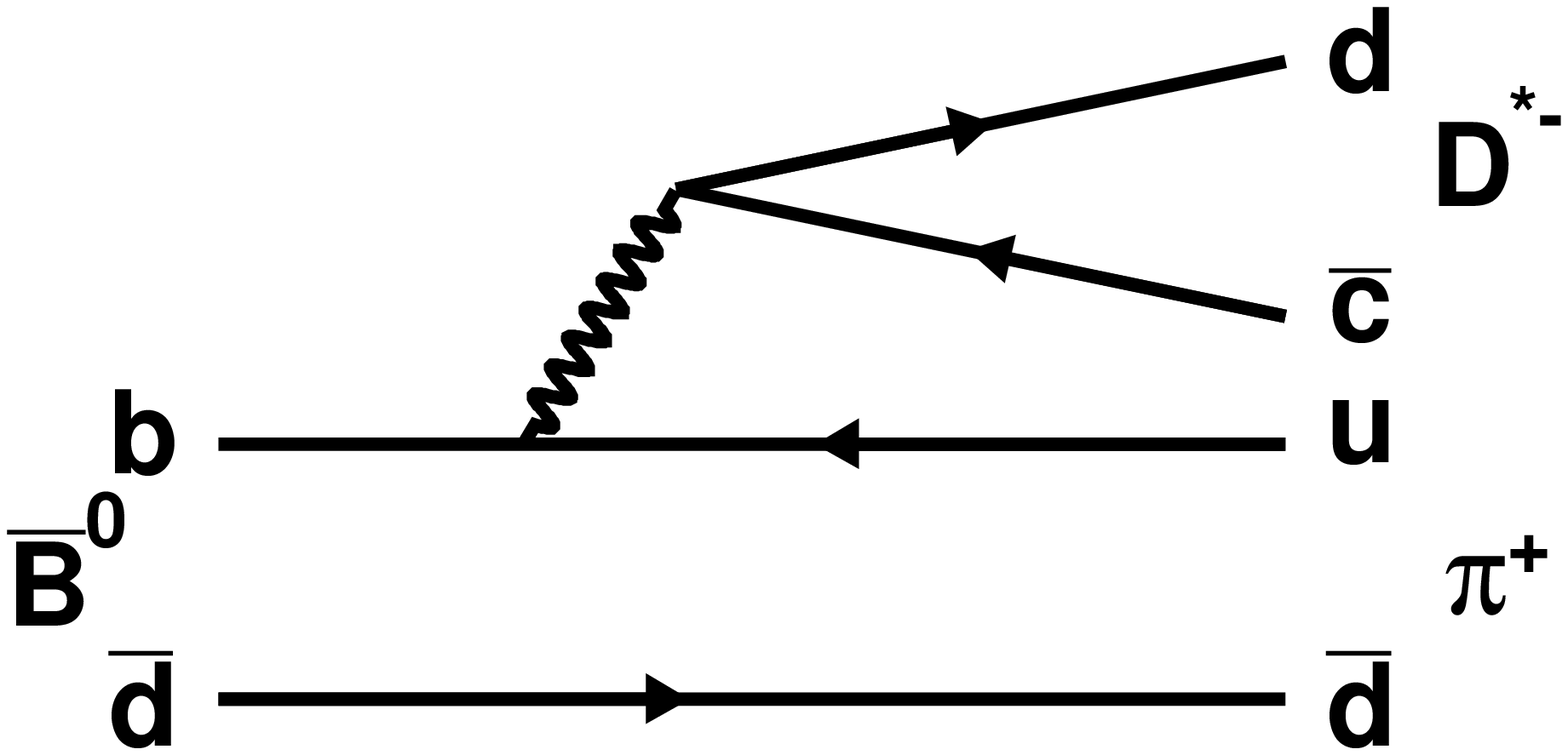}\\
\end{minipage}

    \caption{
      Diagrams for 
       $B^0 \to D^{*-}\pi^+$ (left) and 
      $\overline{B}{}^0 \to D^{*-}\pi^+$ (right).
      Those for $\overline{B}{}^0 \to D^{*+}\pi^-$ and 
      $B^0 \to D^{*+}\pi^-$
      can be obtained by charge conjugation.}
      \label{fig:feynman}
\end{figure}

In the Standard Model (SM), $CP$ violation occurs due to the presence of a complex phase in the Cabibbo-Kobayashi-Maskawa (CKM) matrix~\cite{KM}. Precision measurements of the parameters of the CKM matrix are important to investigate new sources of $CP$ violation. The study of the time-dependent decay rates of $B^0 (\overline{B}{}^0) \to D^{*\mp}\pi^{\pm}$ provides a theoretically clean method for extracting $\sin(2\phi_1+\phi_3)$~\cite{dunietz}, where $\phi_1$ and $\phi_3$ are angles of the CKM Unitarity Triangle as defined in~\cite{pdg}. As shown in Fig.~\ref{fig:feynman}, these decays can be mediated by both Cabibbo-favored (CF) and doubly-Cabibbo-suppressed (DCS) diagrams, whose  
amplitudes are proportional to $V_{cb}^*V_{ud}$ and $V_{ub}^*V_{cd}$, respectively, where $V_{ij}$ are the CKM matrix elements and have a relative weak phase difference $\phi_3$.

The time-dependent decay rates are given by~\cite{fleischer}
\begin{eqnarray}
P(B^{0} &\to& D^{*\pm} \pi^\mp) = \frac{1}{8\tau_{B^0}} 
                    e^{-|\Delta t|/\tau_{B^0}} \nonumber \\ 
&&   \times \left[
      1 \mp C \cos (\Delta m \Delta t) - S^\pm \sin (\Delta m \Delta t) 
    \right],  \nonumber \\
P(\overline{B}{}^0 &\to& D^{*\pm} \pi^\mp) =  \frac{1}{8\tau_{B^0}}  
                  e^{-|\Delta t|/\tau_{B^0}}\nonumber \\ 
&&  \times    \left[
      1 \pm C \cos (\Delta m \Delta t) + S^\pm \sin (\Delta m \Delta t) 
    \right] 
    \label{eq:evol}  
  \end{eqnarray}
  Here $\Delta t$ is the difference between the time of the decay and the 
time that the flavor of the $B$ meson is tagged by the associated $B$ meson,
$\tau_{B^0}$ is the average neutral $B$ meson lifetime, 
$\Delta m$ is the $B^0$-$\overline{B}{}^0$ mixing parameter, and 
$C = \left( 1 - R^2 \right) / \left( 1 + R^2 \right)$, 
where $R$ is the ratio of the magnitudes of the DCS and CF amplitudes
(we assume their magnitudes to be the
same for $B^0$ and $\overline{B}{}^0$ decays). The $CP$ violation parameters for $D^* \pi$ are given by
\begin{equation}
S^{\pm} = \frac{-2 R \sin(2\phi_1+\phi_3 \pm \delta)}
               { \left( 1 + R^2 \right)},
\label{eq:spm}
\end{equation}
where  $\delta$ is 
the strong phase difference between the CF and DCS amplitudes.

 Since the predicted value of $R$ is small, $\sim$0.02~\cite{csr}, we neglect terms of ${\cal O}\left( R^2 \right)$ 
(and hence take $C = 1$). The amount of $CP$ violation in $D^*\pi$ decays, which is proportional to $R$, is expected to be small, and hence, a large data sample is needed in order to
obtain sufficient sensitivity. To increase statistics, we employ a partial 
reconstruction technique~\cite{zheng}, wherein signal is distinguished from background on the basis of 
kinematics of the `fast' pion ($\pi_f$) from the decay $B \to D^* \pi_f$,
and the `slow' pion ($\pi_s$)  from the subsequent decay of  $D^* \to D \pi_s$; thus the $D$ meson is not reconstructed at all. 

     Previous analyses have been reported by Belle~\cite{belle_partial, fronga} as well as by BaBar~\cite{babar_partial}. This study uses a data sample of $605\,\mathrm{fb}^{-1}$ containing 657 x $10^6$ $B \overline{B}$ events. The data sample is about twice the size of the dataset used in the previous Belle analysis ~\cite{fronga} and supersedes the previous study.

The data were collected with the Belle
detector~\cite{Belle} at the KEKB collider~\cite{KEKB} operating near the $\Upsilon$(4S) resonance. 
The Belle detector is a large-solid-angle magnetic 
spectrometer that consists of a silicon vertex detector (SVD), 
a 50-layer central drift chamber (CDC), an array of aerogel 
threshold Cherenkov counters (ACC), a barrel-like arrangement 
of time-of-flight scintillation counters (TOF), and an 
electromagnetic calorimeter (ECL) comprised of CsI(Tl) 
crystals located inside a superconducting solenoidal coil 
that provides a 1.5 T magnetic field. An iron flux-return 
located outside of the coil is instrumented to detect $K_L^0$ 
mesons and to identify muons (KLM). 
A sample containing 
152 x $10^6$ $B \overline{B}$ pairs was collected with 
a 2.0~cm radius beampipe and a 3-layer silicon vertex detector 
(SVD1), 
while a sample of 505 x $10^6$ $B \overline{B}$ pairs was 
collected with a 1.5~cm radius beampipe, a 4-layer silicon vertex
detector (SVD2), and a small-cell inner drift chamber~\cite{svd2}. 

 The ``signal side'' $B$, decaying to  $D^{*+}\pi_f^{-}$, $D^{*+} \to D^{0} \pi_s^{+}$ (or charge conjugate), is reconstructed using pairs of oppositely charged pions. Since the pion originating from the $B$ has a higher momentum in the $\Upsilon(4S)$ c.m. frame than that originating from the $D^{*}$, the former (latter) is referred to as the fast (slow) pion. 
All momenta and energies in this paper are calculated in the $\Upsilon(4S)$ center-of-mass (c.m.) frame, unless otherwise stated.
 Fast pion candidates are required to have a radial (longitudinal) impact parameter $dr <
0.1\,\mathrm{cm}$ ($|dz| < 2.0\,\mathrm{cm}$) and
to have associated hits in the SVD. We reject leptons and kaons based on information from the CDC, TOF and ACC.
A requirement is made on the fast pion momentum, $1.93 \, {\rm GeV}/c < p_{f} < 2.50 \, {\rm GeV}/c$. Soft pion candidates are required to have momenta in the range $0.05 \, {\rm GeV}/c < p_{s} < 0.30 \, {\rm GeV}/c$. No particle  identification requirement is applied for these pions. We impose only a loose requirement that they originate from the run-dependent interaction point (IP) profile. The IP has $\sigma_{z} \sim 4 \,\rm{mm}$ along the beam direction ($z$), and $\sigma_{x} \sim 100 \,\rm{\mu m}$  and $\sigma_{y} \sim 10 \,\rm{\mu m}$ in the plane perpendicular to the beam direction.

For any given $\pi_f$ from a signal $B$ decay, the energy of the $D^*$ may be known through energy conservation, $E_{D^{*}}= E_{B} - E_{\pi_f}$,
where $E_B=\sqrt{s}/2$ at the $\Upsilon$(4S).
The magnitude of the momentum is then 
$|\vec{p}_{D^*}|=\sqrt{ E_{D^{*}}^2 - m_{D^*}^2}$.
Because the $B$ meson is slow in the c.m. frame, its momentum
$|\vec{p}_{B}|=\sqrt{E^2_B - m_{B^0}^2}\approx 0.3\ {\rm GeV}/c$ 
is small relative to the $\pi_f$ and $D^*$ momenta.
It follows from momentum conservation 
\begin{eqnarray}
 \vec{p}_{D^{*}}& = &\vec{p}_{B} -  \vec{p}_{\pi_f}
\label{eqnpr}
 \end{eqnarray} 
that the direction of the $D^*$ momentum can be approximated as the direction opposite to $\vec{p}_{\pi_f}$. 
This approximate $D^*$ four-momentum is denoted as the ``partially reconstructed'' $D^*$.
We define a quantity $p_{\delta}$ = $|{p}{}_{\pi_f}| - |{p}{}_{D^{*}}|$, which for signal decays satisfies 
$|p_{\delta}| \le |\vec{p}_{B}|$, as can be seen by examining Eq.~(\ref{eqnpr}). 

We then examine the soft pion after boosting it into the partially reconstructed $D^{*}$ frame;
in the true $D^{*}$ rest frame, the soft pion is monoenergetic and its momentum has an angular distribution characteristic of a pseudoscalar to pseudoscalar-vector transition, $\propto\cos^2 \theta$ where $\theta$ is taken relative to the boost axis.
In the partially reconstructed frame, 
the momentum will have a limited spread. 
We study the components parallel and perpendicular to the boost axis, denoted  ${p}_{\parallel}$ and ${p}_{\perp}$, respectively. 

We use the three kinematic variables $p_{\delta}$, ${p}_{\parallel}$ and ${p}_{\perp}$ to distinguish between signal and background. Background events are separated into three categories:
$D^{*\mp}\rho^{\pm}$, which is kinematically similar to the signal; 
correlated background, in which the soft pion originates from the decay of 
a $D^*$ that in turn originates from the decay of the same $B$ as the fast pion candidate, excluding $D^{*\mp}\pi^{\pm}$ and $D^{*\mp}\rho^{\pm}$ decays ({\it e.g.}, $B \to D^{**}\pi$, $B \to D^{*}a_{1}$, $B \to D^{*}l\nu$); and uncorrelated background, which includes all other background sources
({\it e.g.}, continuum processes, $B \to D\pi$). The distributions of the kinematic variables for signal and background categories are determined from a large sample of Monte-Carlo (MC) generated data
corresponding to three times the integrated luminosity of our data sample.

  We retain candidates that satisfy $-0.10 \, {\rm GeV}/c < p_\parallel < 0.07 \, {\rm GeV}/c$, $-0.60 \, {\rm GeV}/c < p_\delta < 0.50 \, {\rm GeV}/c$ and ${p}_{\perp} < 0.05$ GeV$/c$.
In the cases where more than one candidate satisfies these criteria,
we select the one with the largest value of $\delta_{\pi_f \pi_s}$, where $\delta_{\pi_f \pi_s}$ is the angle between the fast pion direction and the soft pion direction in the $\Upsilon(4S)$ c.m. frame. The signal region is defined as: $-0.40 \, {\rm GeV}/c < p_\delta <  0.40 \, {\rm GeV}/c$, $-0.05 \, {\rm GeV}/c < p_\parallel < -0.01 \, {\rm GeV}/c$ or $0.01 \, {\rm GeV}/c < p_\parallel  <  0.04 \, {\rm GeV}/c$ and ${p}_{\perp} < 0.05$ GeV$/c$.

The determination of the flavor of the $B$ meson opposite to the signal side $B$, which we refer as the tag-side $B$, is essential for this measurement. In order to tag the flavor of the associated $B$ meson, 
we require the presence of a high-momentum lepton ($l$) in the event. This helps reduce 
background from continuum 
$e^+e^- \to q\overline{q} \ (q = u,d,s,c)$ processes.
Tagging lepton candidates are required to be positively identified
either as electrons, on the basis of information from the CDC, ECL and ACC, 
or as muons, on the basis of information from the CDC and the KLM.
They are required to have momenta in the range
$1.1 \ {\rm GeV}/c < p_{l} < 2.3 \ {\rm GeV}/c$,
and to have an angle with the fast pion candidate that satisfies $\cos \delta_{\pi_f l}>-0.75$ in the $\Upsilon(4S)$ c.m. frame. These requirements reduce to a negligible level (0.7$\%$) the contribution of leptons produced from semileptonic
decays of the unreconstructed $D$ mesons in the $B \to D^{*\mp}\pi^\pm$ decay chain.

  Vertexing requirements identical to those for the fast pion are applied to the lepton candidate in order to obtain an accurate vertex position.
To further suppress the remaining small continuum background,
we impose a loose requirement on the ratio of 
the second to zeroth Fox-Wolfram~\cite{fw} moments, $R_2 < 0.6$.

\begin{figure}[!htb]
\includegraphics[width=9.2cm,clip]{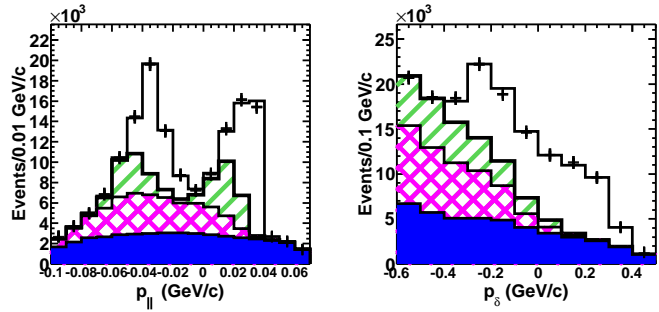}
  \caption{
    \label{fig:kin_fit}
    Results of the fits to $D^*\pi$ candidates projected onto $p_{\parallel}$ (left) and $p_{\delta}$ (right) in the signal region of the two kinematic variables. The contributions are: $D^* \pi$ (open), $D^* \rho$ (green slanted lines), correlated background (magenta crossed lines) and uncorrelated background(shaded blue). Data are shown as points with error bars. 
  }
\end{figure}

Event-by-event signal and background fractions are determined from 
binned maximum likelihood fits to the two-dimensional distributions of $p_\delta$ and $p_\parallel$. 
The results of these fits, projected onto each of the two variables,
are shown in Fig.~\ref{fig:kin_fit},
and summarized in Table~\ref{tab:kin_fit}. We obtain a purity of $59.0\pm0.4\%$ in the signal region, where purity is defined as the ratio of the signal to total yields.

\begin{table}[htb]
   \caption{
     \label{tab:kin_fit}
     Summary of the yields in the signal region
        }
   \begin{center}
     \begin{tabular}{lc}
\hline 
              &    \\
\hline 
$D^* \pi$    & $ 50196\pm286 $  \\
$D^* \rho$   & $  10232\pm 150 $ \\
Correlated background   & $  10425\pm 135 $   \\
Uncorrelated background    & $  14193\pm 128 $  \\
\hline 
     \end{tabular}
   \end{center}
\end{table}

 At the KEKB asymmetric-energy $e^{+}e^{-}$ (3.5 GeV on 8 GeV) collider, operating at the $\Upsilon(4S)$ resonance 
($E_{\rm c.m}$ = 10.58 GeV), the $\Upsilon(4S)$ is produced with a Lorentz boost of $\beta\gamma$ = 0.425, almost
along the electron beamline ($z$). In the $\Upsilon(4S)$  c.m, $B^{0}$ and $\overline{B}{}^0$ mesons are
approximately at rest. Hence the proper time-difference ($\Delta t$) between the  signal side vertex ($z_{\rm sig}$) and the tag-side vertex ($z_{\rm tag}$) is obtained from the fast pion on the signal side and the tagging lepton. The variable $\Delta t$ is defined as:
\begin{eqnarray}
 \Delta t \approx (z_{\rm sig} - z_{\rm tag}) /\beta\gamma c.
\label{eqn2}
\end{eqnarray} 
$z_{\rm sig}$ is obtained from the intersection of the fast pion's track and the IP, and $z_{\rm tag}$ is obtained from the intersection of the tagging lepton's track and the IP.

To measure the $CP$ violation parameters, we perform a simultaneous unbinned fit to four samples:  two are of same-flavor (SF) events, namely $\pi^{+}l^{+}$, $\pi^{-}l^{-}$, in which the fast pion and the tagging lepton
have the same charge, and the other two are of opposite-flavor (OF) events, namely $\pi^{+}l^{-}$, $\pi^{-}l^{+}$, in which the
fast pion and the tagging lepton have opposite charge. We minimize the quantity $-\ln {\cal L} = - \sum_i \ln {\cal L}_i$,
where 
\begin{equation}
  \label{eq:likelihood}
  {\cal L}_i = 
  f_{D^*\pi} P_{D^*\pi} + f_{D^*\rho} P_{D^*\rho} +
  f_{\rm unco} P_{\rm unco} + f_{\rm corr} P_{\rm corr}.
\end{equation}
Here, $f_{x}$ stands for the event-by-event fraction from source $x$
and is obtained from the fits to the kinematic variables, and $P$ denotes the probability density functions (PDFs) for signal and backgrounds, which contain an underlying physics PDF with experimental effects taken into account. The convolution of the physics PDF with experimental effects will be described later. For $D^*\pi$ and $D^*\rho$, the PDF is given by Eq.~(\ref{eq:evol}),
whereas for $D^*\rho$ the $S^\pm$ terms are effective parameters
averaged over the helicity states~\cite{dstarrho} and are constrained to be zero.
The PDF for correlated background contains
a term for neutral $B$ decays 
(given by Eq.~(\ref{eq:evol}) with $S^\pm = 0$),
and a term for charged $B$ decays
(for which the PDF is 
$\frac{1}{2\tau_{B^+}} e^{-\left| \Delta t \right| / \tau_{B^+}}$,
where $\tau_{B^+}$ is the lifetime of the charged $B$ meson). The PDF for uncorrelated background also contains
neutral and charged $B$ components, with the remainder from continuum 
$e^+e^- \to q\overline{q} \ (q = u,d,s,c)$ processes.
The continuum PDF is modeled with two components: 
one with negligible lifetime,  and the other with a finite lifetime, which takes into account the dependence of average lifetime of the charm contribution in the continuum (close to the average $D$ meson lifetime).

The parameters in $P_{\rm unco}$ and $P_{\rm corr}$ are obtained from separate simultaneous fits to OF and SF candidates in the respective sideband regions, defined later. Since there is no $CP$ violation in background, the corresponding parameters are fixed to zero in these fits. The fit is further simplified by fixing the biases in $\Delta z$ to zero (discussed later in detail).  MC simulation studies demonstrate that varying or fixing these biases to zero does not affect the background parameters.
  
\begin{figure}[!htb]
\includegraphics[width=6.2cm,clip]{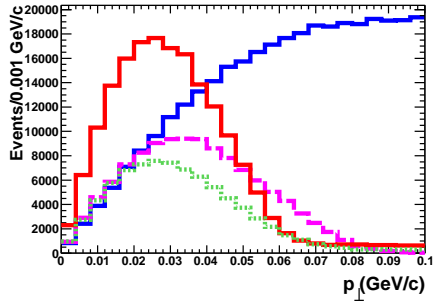}
  \caption{
    \label{fig:p_perp}
     $p_{\perp}$ distributions for various MC simulations, showing the following contributions: $D^* \pi$ (red solid line), $D^* \rho$ (dotted green line), correlated background (dashed magenta line), and uncorrelated background (blue solid line).
  }
\end{figure}

To measure the uncorrelated background shape, we use events in a sideband region, $-0.10 \, {\rm GeV}/c < p_{\parallel} < -0.07 \, {\rm GeV}/c$ or $0.01\, {\rm GeV}/c<p_{\parallel} <0.04\, {\rm GeV}/c$, $-0.60 \, {\rm GeV}/c < p_{\delta} < 0.50 \, {\rm GeV}/c$ and $0.08 \, {\rm GeV}/c <p_{\perp} < 0.10 \, {\rm GeV}/c$, which is populated mostly by uncorrelated background ($\sim 90 \%$). To determine the correlated background parameters, we use events 
in a sideband region,
$-0.10 \, {\rm GeV}/c < p_{\parallel} < -0.07 \, {\rm GeV}/c$,
$-0.60 \, {\rm GeV}/c < p_{\delta}< 0.00 \, {\rm GeV}/c$ and
$0.00 \, {\rm GeV}/c < p_{\perp} < 0.05 \, {\rm GeV}/c$. This sideband region is dominated by both correlated and uncorrelated backgrounds and has a very small amount of $D^*\pi$ signal and $D^*\rho$ background. The uncorrelated background parameters are fixed to the values obtained in the previous fit. Figure~\ref{fig:p_perp} shows $p_{\perp}$ distributions for signal and various background components in MC simulations, corresponding to about three times the size of the data.

    The PDF for signal and background in Eq.~(\ref{eq:likelihood})
must be convolved with the corresponding $\Delta z$ resolution functions related to the 
kinematic smearing (${\mathcal{R}_{\rm k}}$), detector resolution (${\mathcal{R}_{\rm det}}$), and asymmetry in $\Delta z$ from non-primary tracks (${\mathcal{R}_{\rm np}}$). The resolution function related to kinematic smearing is due to the fact that we use the approximation of Eq.~(\ref{eqn2}). The detector resolution and smearing due to the asymmetry in $\Delta z$ from non-primary tracks are described in detail elsewhere~\cite{fronga}.

To account for mistagging, the PDFs in Eq.~(\ref{eq:likelihood}) are divided into two components
\begin{eqnarray}
    \label{eq:exp_pdf}
   P( l^{\mp}, \pi_f^\pm)  =    ( 1 - w_{\mp} )  P(B^{0} /\overline{B}{}^0 \to D^{*\mp} \pi^\pm)\nonumber \\
  + 
w_{\pm} P(\overline{B}{}^0/ B^{0} \to D^{*\mp} \pi^\pm),
 \end{eqnarray}
where $w^{+}$ and $w^{-}$ are the wrong-tag fractions, defined as the probabilities to incorrectly measure the flavor of tagged $B^{0}$ and $\overline{B}{}^0$ mesons, respectively, and are determined from the data as free parameters in the fit for $S^\pm$.

The time difference $\Delta t$ is related to the measured quantity $\Delta z$
as described in Eq.~(\ref{eqn2}), with an additional term due to 
possible offsets in the mean value of $\Delta z$,
\begin{equation}
  \label{eq:dt_offset}
  \Delta t \longrightarrow \Delta t + \epsilon_{\Delta t} \simeq \left( \Delta z + \epsilon_{\Delta z} \right) / \beta\gamma c.
\end{equation}
It is essential to allow non-zero values of $\epsilon_{\Delta t}$ since a 
small bias can mimic the effect of $CP$ violation:
\begin{equation}
      \cos (\Delta m \Delta t) 
 \to
      \cos (\Delta m \Delta t) - 
\Delta m \epsilon_{\Delta t} \sin (\Delta m \Delta t)
\end{equation}
A bias as small as $\epsilon_{\Delta z} \sim 1 \ \mu{\rm m}$ can lead to 
sine-like terms as large as $0.01$, 
comparable to the expected size of the $CP$ violation effect.
 Because both vertex positions are obtained from single tracks,  
the partial reconstruction analysis is more susceptible than other Belle $CP$ 
violation analyses to 
such biases. We allow separate offsets for $\Delta z$ for each combination of $\pi_{h}$ and $l$ charges.   Thus we have eight offsets in total, four for each data sample, SVD1 and SVD2.

\begin{figure}[!htb]
  \begin{center}
    \includegraphics[width=9.2cm]{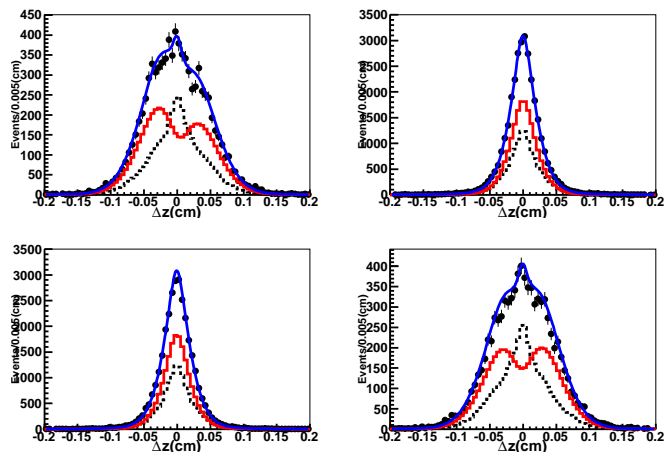}

  \end{center}
  \caption{
    \label{fig:myfit}
    $\Delta z$ distributions for four flavor-charge combinations: $\pi^{-}l^{-}$ (top left) , $\pi^{-}l^{+}$ (top right), $\pi^{+}l^{-}$ (bottom left), and $\pi^{+}l^{+}$ (bottom right). The fit result (solid blue line) is superimposed on the data (solid points with error-bars).  The signal and background components are shown as the solid red and dotted black curves, respectively.
   }
\end{figure}

To extract the $CP$ violation parameters we fix $\tau_{B^0}$ and 
$\Delta m$ at their world average values ($\tau_{B^0} = 1.530 \pm 0.009 \ {\rm ps}$
and $\Delta m = 0.507 \pm 0.005 \ {\rm ps}^{-1}$~\cite{pdg}),
and fit with $S^+$, $S^-$, two wrong tag fractions, and eight offsets
as free parameters. 
We obtain
$S^+ =  +0.061\pm0.018$ and $S^- =  +0.031\pm0.019$, where the errors are statistical only.
The wrong tag fractions are $w_- = (5.3 \pm 0.3)\%$ and 
$w_+ = (5.2 \pm 0.3)\%$. 
All floating offsets are consistent with zero except  for one of the OF combinations ($\pi_{f} = \pi^{-}$, $l = \ell^{+}$) in the SVD1 sample.
The results are shown in Fig.~\ref{fig:myfit}. Using large MC samples generated with non-zero and zero $S^\pm$ values, we do not find any significant bias in the procedure.

\begin{figure}[!htb]
  \begin{center}
    \includegraphics[width=9.2cm]{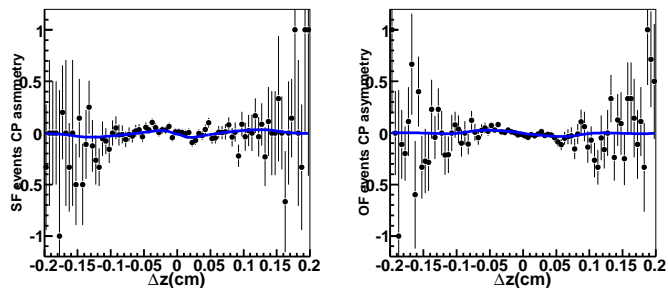}
  \end{center}
  \caption{
    \label{fig:myfit_asp}
    Results of the fit to obtain $S^+$ and $S^-$,
    shown as asymmetries in the SF events (left)  and OF events (right).
    The fit results (solid blue lines) are superimposed on the data.
  }
\end{figure}

  To further illustrate the $CP$ violation effect,
we define asymmetries in the 
same flavor events (${\cal A}^{\mathrm{SF}}$) 
and in the opposite flavor events (${\cal A}^{\mathrm{OF}}$), as 
\begin{eqnarray}
  {\cal A}^{\mathrm{SF}} & = &
  \frac{ N_{\pi^-l^-}(\Delta z) - N_{\pi^+l^+}(\Delta z) } 
       { N_{\pi^-l^-}(\Delta z) + N_{\pi^+l^+}(\Delta z) }, \nonumber \\
  {\cal A}^{\mathrm{OF}} & = &
  \frac{ N_{\pi^+l^-}(\Delta z) - N_{\pi^-l^+}(\Delta z) } 
       { N_{\pi^+l^-}(\Delta z) + N_{\pi^-l^+}(\Delta z) }, 
\end{eqnarray}
where the $N$ values denote the number of events 
for each combination of $f$ and $l$ charge.
These are shown in Fig.~\ref{fig:myfit_asp}.

This analysis is very sensitive to the vertexing bias.  Hence, we include $\Delta z$ offsets in the fits to account for this bias. In order to estimate the error due to these offsets, we perform fits to obtain $S^\pm$ values with and without offsets using an ensemble of 100 generated $D^{*}\pi$ signal samples, and use the difference between the two results as the systematic error. We obtain negligible contribution to the systematic errors when we float $\Delta z$ offsets in the background PDF.

  Other sources of systematic error are the resolution functions, ${\mathcal{R}_{\rm k}}$, ${\mathcal{R}_{\rm det}}$ and ${\mathcal{R}_{\rm np}}$, uncorrelated and correlated backgrounds and physics parameters, $ \Delta m$, $ \tau_{B^0}$ and $ \tau_{B^+}$ that are fixed in the fit to extract $S^{\pm}$. The parameters of the resolution functions and backgrounds are varied by $\pm 1 \sigma$ (with $\Delta m$ and $\tau_{B^0}$ fixed), respectively, where $\sigma$'s are the corresponding errors of the parameters and the difference is assigned as systematic error. We vary the physics parameters by $\pm 1 \sigma$, where $ \sigma$ is the error of the corresponding PDG values, and we then use the difference between the $S^{\pm}$ values thus obtained and the default values as the systematic error. When the fit is performed floating $S^\pm$ values, along with $\tau_{B^0}$ and $\Delta m$, we obtain:  $S^+ =  +0.055\pm0.018$ and $S^- =  +0.039\pm0.019$, $\tau_{B^0} = 1.550 \pm 0.008 \ {\rm ps}$ and $\Delta m = 0.473 \pm 0.004 \ {\rm ps}^{-1}$, where the errors are statistical only. The deviations from the nominal fit (0.06, 0.08) are close to the systematic errors assigned for the physics parameters (Table~\ref{tab:systematics}). The difference between the $S^{\pm}$ values obtained floating both $\Delta m$ and $\tau_{B^0}$ parameters and the default value is also added to the systematic error estimation. In the fits to extract $S^{\pm}$, ${S^{\pm}}_{D^{*}\rho} $ and ${S^{\pm}}_{\rm corr} $ are set to zero. For the systematic error due to these parameters, the fit is performed with these values set to $\pm 0.05$ and the difference between the $S^{\pm}$ value thus obtained and the default value is assigned as the systematic error.

We use a triple Gaussian to model the detector resolution ($R_{\rm det}$) function. We consider the systematic uncertainty due to the lack of knowledge of the exact functional form of the resolution model.  When the resolution models are varied, we obtain shifts as large as $0.006$ for $S^{+}$. This is conservatively assigned as the systematic error due to this source. 
 
 We obtain a vertexing systematic error of 0.003 for $S^{\pm}$. Additional systematic errors result from varying the number of bins for the kinematic variables, $p_{\delta}$ and ${p}_{\parallel}$ in the yield fit.

The systematic errors are summarized in Table~\ref{tab:systematics}.
The total systematic error is obtained by adding the 
above terms in quadrature.
\begin{table}[htb]
 \caption{Summary of possible sources of systematic error }
\label{tab:systematics}
  \begin{center}
   \begin{tabular}{lccc} \hline  
  Systematic error source& $ S^{+}$ &$S^{-}$ \\\hline 
  $\Delta z$ offset &$0.002$&$0.003$\\
   ${\mathcal{R}_{k}}$ parameters  &$0.002$&$0.003$\\
  ${\mathcal{R}_{det}}$ parameters&$0.002$&$0.002$\\
 ${\mathcal{R}_{np}}$ parameters &$0.004$&$0.004$\\
 Background parameters &$0.001$&$0.001$\\
 Physics parameters &$0.006$&$0.009$\\
  Floating $\tau_{B^0}$ and $\Delta m$ &$0.006$&$0.008$\\
 Yield fit &$0.003$&$0.005$\\
 Resolution model &$0.006$&$0.002$\\
IP constraint&$0.003$&$0.003$\\\hline
 Total systematic error &$0.012$&$0.015$\\
    \hline 
    \end{tabular}
   \end{center}
\end{table}
 
    In conclusion, we have measured $CP$ violation parameters that depend on $\phi_3$ using the 
time-dependent decay rates of $B^0 \to D^{*\mp} \pi^\pm$ with a data sample containing 657 x $10^6$ $B \overline{B}$ events. We determine the $CP$ violation parameters $S^{\pm}$ to be
\abovedisplayskip=6pt
\belowdisplayskip=6pt
\begin{eqnarray}
  S^+ & = &  +0.061\pm 0.018 \pm 0.012, \nonumber \\ 
  S^- & = & +0.031 \pm 0.019 \pm 0.015, 
\end{eqnarray}
where the first errors are statistical and the second errors are systematic. We can also express the results as parameters $a$, $c$, defined as:
\begin{eqnarray}
a = - (S^+ + S^-)/2,\nonumber \\
c = - (S^+ - S^-)/2.
\label{eqn5}
\end{eqnarray} 
Our results thus become:
\begin{eqnarray}
 a & = &  -0.046\pm 0.011 \pm 0.015 ,  \nonumber \\
 c & = &  -0.015 \pm 0.011 \pm 0.015.
\end{eqnarray}
\setlength{\parskip}{\baselineskip}
      The deviation of $a$ from zero is a measure of the amount of $CP$ violation. We obtain a significance of 2.5$\sigma$ on the $CP$ violation parameter, $a$. Our measurement is consistent with the world average value and significantly improves the precision of previous measurements reported by Belle~\cite{belle_partial, fronga} as well as by BaBar~\cite{babar_partial} and supersedes our earlier result~\cite{fronga}.
      
        We thank the KEKB group for excellent operation of the accelerator, the KEK cryogenics group for efficient solenoid operations, and the KEK computer group and the NII for valuable computing and SINET3 network support. We acknowledge support from MEXT, JSPS and Nagoya's TLPRC (Japan);
ARC and DIISR (Australia); NSFC (China); MSMT (Czechia); DST (India); MEST, NRF, NSDC of KISTI, and WCU (Korea); MNiSW (Poland); MES and RFAAE (Russia); ARRS (Slovenia); SNSF (Switzerland); 
NSC and MOE (Taiwan); and DOE (USA).

\end{document}